\documentclass[runningheads]{llncs}

\usepackage{cite}
\usepackage{amsmath,amssymb,amsfonts}
\usepackage{algorithmic}
\usepackage{graphicx}
\usepackage{textcomp}
\usepackage{xcolor}
\def\BibTeX{{\rm B\kern-.05em{\sc i\kern-.025em b}\kern-.08em
    T\kern-.1667em\lower.7ex\hbox{E}\kern-.125emX}}

\usepackage{latexsym, bm, bbm, nicefrac, stmaryrd, enumitem, centernot}
\usepackage{url, rotating}
\definecolor{svlinks}{rgb}{.0,0.3,0.6}
\usepackage[hang]{footmisc}

\usepackage[T2A,T1]{fontenc}
\usepackage[utf8]{inputenc}
\usepackage[russian,greek,UKenglish]{babel}

\usepackage{epsfig, listings}
\usepackage{multirow}
\usepackage{float}
\usepackage[section]{placeins}
\usepackage[all]{nowidow}
\usepackage{fnpct}
\usepackage{soul}

\usepackage{tikz, wrapfig}
\usetikzlibrary{matrix}
\usepackage[normalem]{ulem}
\usepackage{booktabs}
\usepackage{array}
\usepackage[subfolder]{gnuplottex}
\usepackage{xfrac}
\usepackage{float}

\usepackage{placeins}
\usepackage[all]{nowidow}
\usepackage{fnpct}
\usepackage{orcidlink}
\usepackage[capitalise,noabbrev]{cleveref}
\usepackage{caption}

\setlist{nosep}

\usepackage{etoolbox}
\pretocmd{\eqref}{Eq.~}{}{}

\hypersetup{
pdftitle={Towards Measuring Vulnerabilities and Exposures in Open-Source Packages},
pdfsubject={},
pdfauthor={Tobias Dam, Sebastian Neumaier},
pdfkeywords={open source, vulnerabilities, CVE, software repository}
}

\begin{document}
\title{Towards Measuring Vulnerabilities and Exposures in Open-Source Packages}
%
%
\author{Tobias Dam\orcidID{0000-0002-2463-5831} \and
Sebastian Neumaier\orcidID{0000-0002-9804-4882}}

\authorrunning{T. Dam and S. Neumaier}
%
\institute{St.\ P\"olten University of Applied Sciences, Austria\\
\email{firstname.lastname@fhstp.ac.at}}
\maketitle              
\begin{abstract}
Much of the current software depends on open-source components, which in turn have complex dependencies on other open-source libraries. Vulnerabilities in open source therefore have potentially huge impacts. The goal of this work is to get a quantitative overview of the frequency and evolution of existing vulnerabilities in popular software repositories and package managers. To this end, we provide an up-to-date overview of the open source landscape and its most popular package managers, we discuss approaches to map entries of the Common Vulnerabilities and Exposures (CVE) list to open-source libraries and we show the frequency and distribution of existing CVE entries with respect to popular programming languages.
\end{abstract}

\section{Introduction}




According to a 2021 open-source security report by Synopsis,\footnote{\url{https://www.synopsys.com/software-integrity/resources/analyst-reports/open-source-security-risk-analysis.html}, last accessed 2023-02-09} 98\% of 1,5k reviewed codebases depend on open-source components and libraries. Given the number of dependencies of medium- to large-size software projects, any vulnerability in open-source code has security implications in numerous software products and involves the risk of disclosing vulnerabilities either directly or through dependencies, as famously seen in the 2014 Heartbleed Bug\footnote{\url{https://heartbleed.com/}, last accessed 2022-12-06}, a vulnerability in OpenSSL which exposed large parts of the existing websites at this time.

Open-source code is written in various programming languages and published in corresponding package managers. 
Currently, the largest package managers are the platform of the Go programming language and the NPM repository for Node.js, cf. \Cref{tab:reponumbers}. 
The documentation and communication of discovered vulnerabilities, however, does not take place directly at the package managers.
The most important platform for discovered vulnerabilities is the Common Vulnerabilities and Exposures (CVE) list \cite{mann1999towards}.
\footnote{\url{http://cve.mitre.org}, last accessed 2023-02-09} It is a dictionary and reference of common and publicly known IT system vulnerabilities, operated by the Mitre Corporation, an American non-profit institution. 
CVE is a critical source for security management, however, is to some extend an heterogeneous and unstructured source. 
In particular, it does not include structured pointers and references to package managers (e.g., NPM) and/or the source code in software repositories (e.g., Github). 
Therefore, CVE does not provide enough information to get an overview of the status and evolution of existing vulnerabilities in the different software repositories and package managers.

The goal of this work is to study the open research problem of mapping CVE entries to open-source projects. To do so, we focus on the following contributions:
\begin{itemize}
    \item We provide an overview of the landscape of open-source projects and libraries in popular package managers. 
    \item We discuss and implement three concrete approaches to map CVE entries to open-source projects.
    \item We perform a frequency analysis of CVE entries corresponding to open-source packages based on the mapping approaches.
    \item Eventually, we discuss identified challenges and quality issues wrt. the available data and the mapping.
\end{itemize}	


\section{Related Work}
\label{sec:relatedwork}


The overview and analysis of the open-source landscape in this paper is based on Libraries.io \cite{jeremy_katz_2020_3626071}. Alternatively, there are various other works that monitor existing repositories and provide quality and popularity metrics. For instance, the PyDriller framework \cite{Spadini2018} is a tool to mine git-based software repositories. In this respect, the GHTorrent framework \cite{gousios2013ghtorent} collects and provides data for all public projects available on Github and therefore is a very comprehensive resource for analysis. 

The detection and reporting of vulnerabilities and threats in open-source software has been the subject of extensive research for several years already \cite{DBLP:conf/ccs/EdwardsC12,Kaplan2021,DBLP:conf/msr/DecanMC18}. 
In a recent paper, Tan et al. report the deployment of security patches on stable branches of open-source projects \cite{10.1145/3485447.3512236}. Similar to our approach, the authors map CVE entries based on the name of an open-source package. The approach is however based on manual investigation of the software packages and the mappings. 

Most related to our mapping of CVE entries to libraries in open-source repositories is the work of Snyk.io\footnote{\url{https://snyk.io/}, last accessed 2023-02-09} which provides a monitoring service to identify vulnerable packages and libraries. 
It consists of a detailed list of security reports for various repositories and contains information about the vulnerability, the name of the affected library, the versions of the library that are affected, etc. 
Similarly, the GitHub Advisory Database\footnote{\url{https://github.com/advisories}, last accessed 2023-02-09}  is a list of CVEs and GitHub originated security reports and vulnerabilities of software available on the GitHub platform.

\section{Open Source Landscape\label{sub:open_source_packages}}

In the context of this paper, open source refers to source code that is made available in online software repositories for use or modification. 
In this work, we make use of the monitoring project \textit{Libraries.io} \cite{jeremy_katz_2020_3626071}: it monitors and crawls the meta-information of over 30 package managers and indexes data from over 5 million projects. Libraries.io provides the collected information as open data; in this paper we make use of the January 2020 dump of the Libraries.io data which is hosted on Zenodo \cite{jeremy_katz_2020_3626071}.

\begin{table}[!htb]
    \caption{The left table lists the number of projects in the top-7 package managers and the middle table the top-7 licenses across all packages. The right table gives the number of versions published at repositories in 2015 and 2019, respectively.\label{tab:reponumbers}}
    \begin{minipage}[t]{.33\linewidth}
      \centering
    \begin{tabular}[t]{lrr}
    \toprule
      P.M. &  Projects &  Perc. \\
    \midrule
            Go &  1,818,666 & 39.45 \\
           NPM &  1,275,082 & 27.66 \\
     Packagist &   313,278 &  6.80 \\
          Pypi &   231,690 &  5.03 \\
         NuGet &   199,447 &  4.33 \\
         Maven &   184,871 &  4.01 \\
      Ruby &   161,608 &  3.51 \\
    \midrule
        \textit{Others} &   425,020 &  9.22 \\
    \bottomrule
    \end{tabular}
    \end{minipage}%
    \begin{minipage}[t]{.33\linewidth}
      \centering
\begin{tabular}[t]{lrr}
\toprule
      License &    Count &  Perc. \\
\midrule
          MIT &  1,637,451 & 44.13 \\
   Apache-2.0 &   848,475 & 22.87 \\
          ISC &   332,676 &  8.97 \\
        Other &   298,626 &  8.05 \\
 BSD-3        &   140,806 &  3.80 \\
      GPL-3.0 &    63,890 &  1.72 \\
      MPL-2.0 &    51,517 &  1.39 \\
    \midrule
       \textit{Others} &   336822 &  9.08 \\
\bottomrule
\end{tabular}
    \end{minipage} 
    \begin{minipage}[t]{.33\linewidth}
      \centering
    \begin{tabular}[t]{lrr}
    \toprule
      P.M. & 2015 & 2019 \\
    \midrule
           NPM & 716,207 & 3,892,909 \\
         NuGet & 210,929 & 614,774 \\
          Pypi & 150,591 & 469,753 \\
     Packagist & 233,643 & 334,059 \\
         Maven & 326,089 & 144,813 \\
    \midrule
        \textit{Others} & 459,265 & 364,164 \\
    \bottomrule
    \end{tabular}
    \end{minipage} 
\end{table}


As shown in \Cref{tab:reponumbers}, currently the largest package managers are the platforms of the Go programming language and the Node.js platform NPM. According to the 2020 data, the Go platform makes up even 39\% of all projects indexed in Libraries.io by listing 1.8 million projects; NPM lists around 1.3 projects which amount to 28\%. The \textit{Others} category in the left table includes 25 other repositories which only sum up to 9\% of all indexed projects. Note that the size of a package managers (i.e. the number of versions hosted) does not necessarily correlate with the impact of these managers. For instance, while the language Go has the most populous package manager, it has not been in the 10 most popular languages in the last years.\footnote{\url{https://octoverse.github.com/2022/top-programming-languages}, last accessed 2023-03-20}

The middle table lists the 7 most common licenses across all packages and repositories. 
The most popular license is the MIT License (44\% of the projects), which is a permissive license, i.e. it imposes only very limited restrictions on reuse and is highly compatible with other open-source licenses.

\begin{figure}
  \begin{minipage}{0.5\textwidth}
        \centering
        \includegraphics[width=1\linewidth]{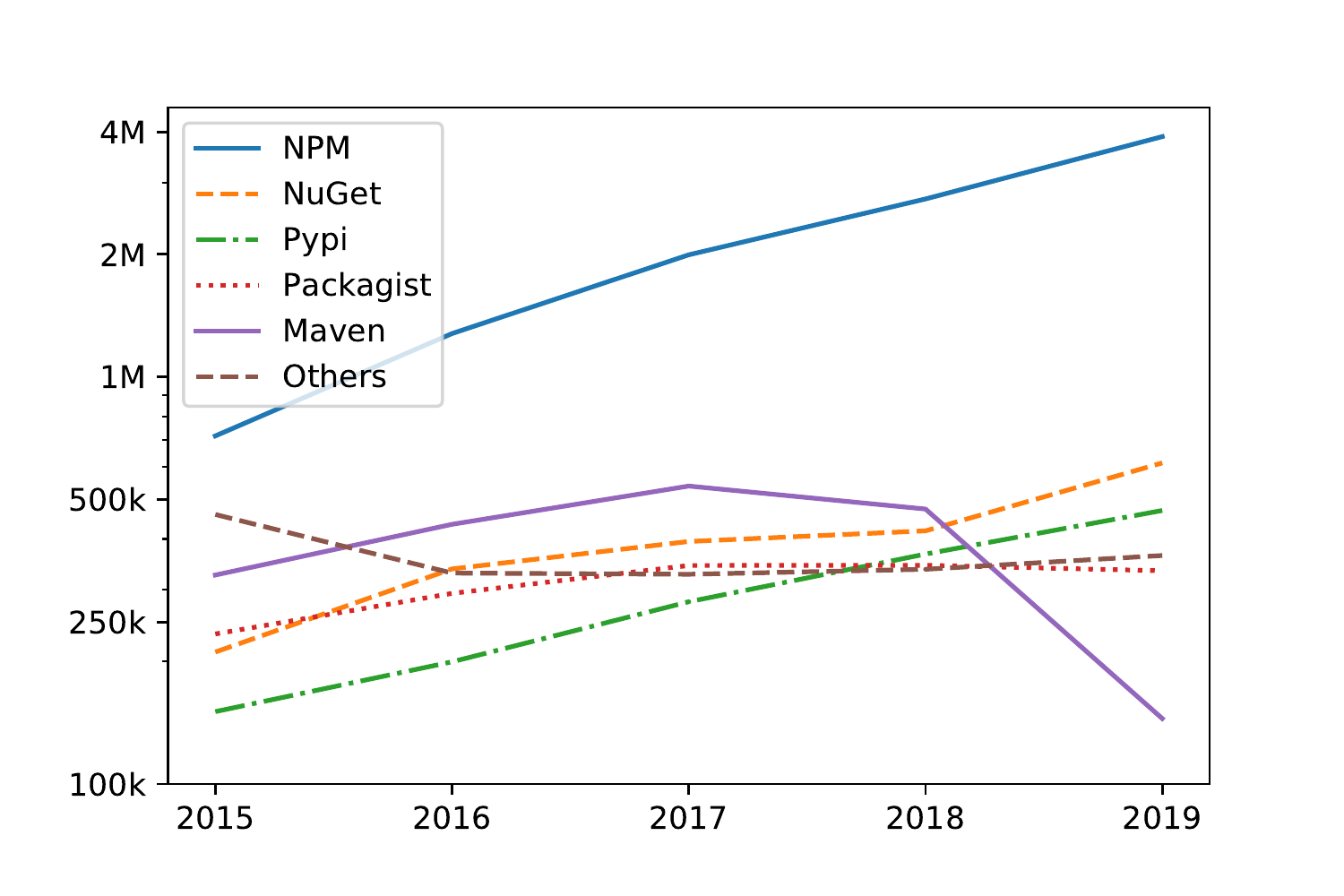}
        \caption{Evolution of the total number of published versions at top-5 largest repositories; the y-axis is in logarithmic scale.}
        \label{fig:versions_evolution}
  \end{minipage}
  \hfill
  \begin{minipage}{0.48\textwidth}
    \centering
    \begin{tabular}{rr}
    \toprule
    Year & CVE entries \\
    \midrule
    2015 & 6,602  \\
    2016 & 6,520  \\
    2017 & 18,161 \\
    2018 & 18,213 \\
    2019 & 19,095 \\
    2020 & 20,516 \\
    2021 & 22,226 \\
    \bottomrule
    \end{tabular} 
    \vspace{.6cm}
    \captionof{table}{Number of new CVE entries in recent years.\label{tab:cve_evolution}}
    \end{minipage}
\end{figure}

A version in the Libraries.io data corresponds to an immutable published version of a project from a repository. 
\Cref{tab:reponumbers} (right) displays the number of versions published at the monitored repositories in 2015 and 2019, respectively. The largest repository NPM increased considerably from 716k to 3.9 million published versions per year. 
\Cref{fig:versions_evolution} visualizes the rapid growth, in number of published versions, of the package managers NPM, NuGet, and PyPI.

\vspace{-0.5cm}

\section{Common Vulnerabilities and Exposures\label{ssec:cve}}

In 1999, Mann and Christey~\cite{mann1999towards} proposed the Common Vulnerability and Exposures (CVE) List, a public list of exposure references. The goal was to find out and document if multiple tools had identified the same vulnerabilities. 

Each \textit{CVE Identifier} in the list identifies a concrete vulnerability and gives an unique, individual name to it. New CVE identification numbers are assigned by the CVE Numbering Authorities (e.g., software vendors, open-source projects, security researchers).
\Cref{tab:cve_evolution} displays the number of newly published CVE entries in the last 7 years. This number increased from 6.6k to 22.2k new entries.


Besides the ID and the summary of the vulnerability, a CVE entry provides relevant references (e.g., link to Github documentation), an assessment of the severity of the vulnerability using the Common Vulnerability Scoring System (CVSS)
, and the Common Weakness Enumeration (CWE) taxonomy of weakness types. 
Additionally, the CVE entry provides Common Platform Enumerations (CPE)~\cite{cheikes2011common} version 2.3 string, which is a standard for identifying classes of applications, operating systems/platforms, as well as hardware information. 
The CPE string consists of various fields providing information such as the software name, vendor as well as the target software, which contains the software computing environment (e.g. ``node.js'' or ``python'').



\section{Findings}
\label{sec:evaluation}




To perform this study, we use the most recent Open Data dump of the Libraries.io database\footnote{\url{https://zenodo.org/record/3626071/files/libraries-1.6.0-2020-01-12.tar.gz}, last accessed 2023-02-09} -- available as compressed archive of several CSV files -- to obtain data about open-source repositories and respective libraries; details of the repositories are described in \Cref{sub:open_source_packages}. The Libraries.io data includes the fields \textsc{Name}, \textsc{Platform} (the package manager), \textsc{ID}, \textsc{Repository Type} (e.g., GitHub), \textsc{Repository Link}, \textsc{Repository Owner} and \textsc{Keywords}.

We used the dockerized version of the open-source project CVE-Search\footnote{\url{https://github.com/cve-search/cve-search}, last accessed 2023-02-09} as basis for our CVE analyses. 
The Computer Incident Response Center Luxembourg (CIRCL) operates a publicly accessible instance of the CVE-Search project and provides a daily dump of their CVE data in JSON format\footnote{\url{https://www.circl.lu/opendata/}, last accessed 2023-02-09}, which we imported into our database. 
The CIRCLE CVE data provides the CVE fields \textsc{Summary} and \textsc{References}, as well as the fields \textsc{Product} and \textsc{Target Software} which hold extracted information from the Common Platform Enumerations (CPE) information.

\subsection{Mapping Approaches}

In order to establish a mapping between the software packages of the Libraries.io dataset and the corresponding CVE entries of the CIRCL CVE dataset, we applied three different approaches, yielding varying results: 

\subsubsection{Strict Name Mapping\label{ssec:strict}}
The first approach iterates over each entry contained in the Libraries.io dataset and checks whether the package manager (in the field \textsc{Platform}) can be matched to any target software in the CIRCL CVE dataset (as stated in the \textsc{Target Software} field).

For our query, we created a mapping between package manager values of the Libraries.io dataset and the corresponding \textsc{Target Software} values. The query then matches the value of the Libraries.io \textsc{Name} field with the values of the CVE \textsc{Product} field as well as the \textsc{Platform} value with either the \textsc{Target Software} or checks if it is contained in the \textsc{Summary} of the CVE dataset.

\begin{figure}[t]
    \centering
    \begin{minipage}[t]{.49\linewidth}
      \centering
    \includegraphics[width=\linewidth]{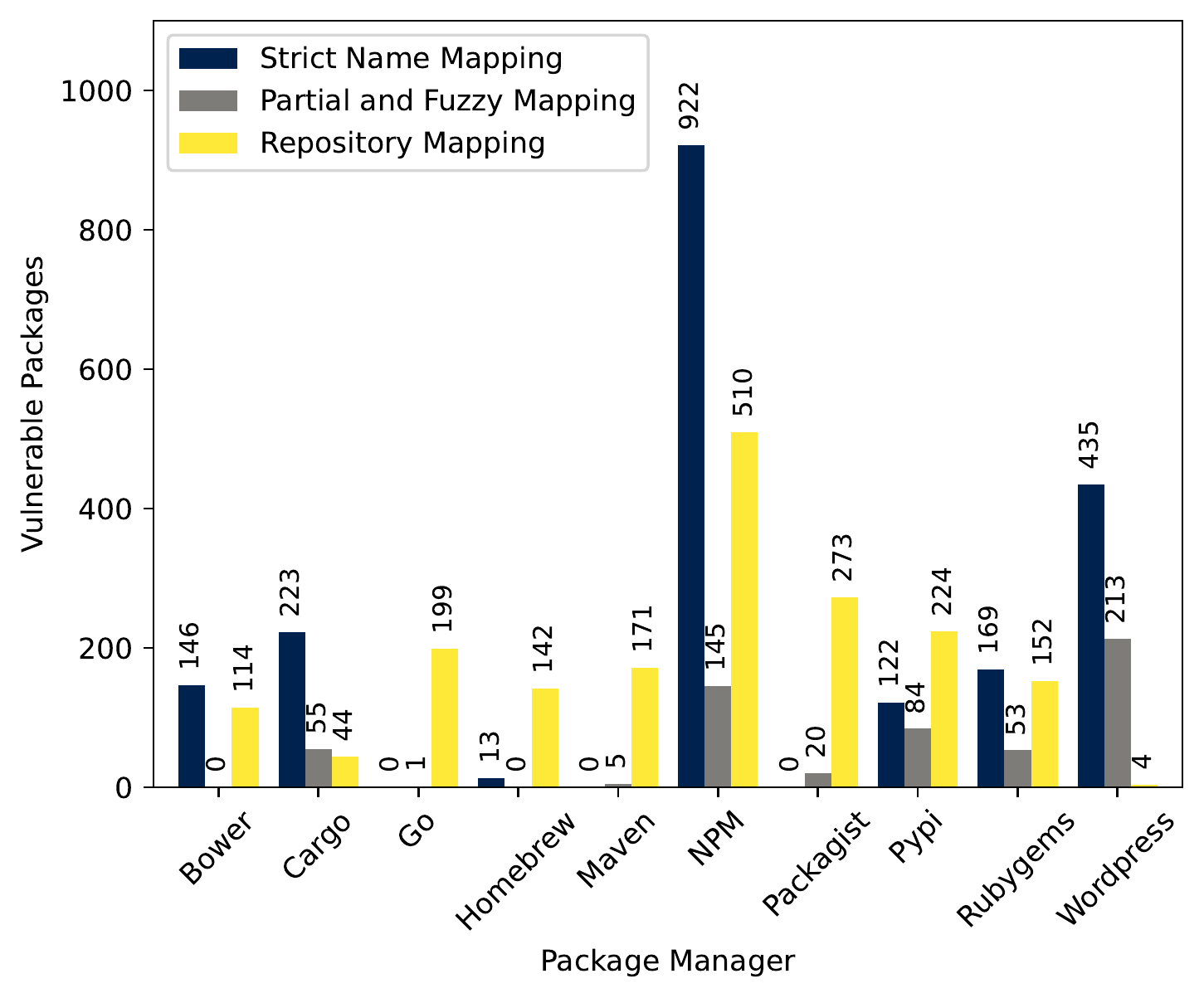}
    \caption{Vulnerable package count of the top-10 package manager across all three mapping approaches.\label{fig:cve_per_pm}}
    \end{minipage}
    \begin{minipage}[t]{.49\linewidth}
        \centering
    \includegraphics[width=\linewidth]{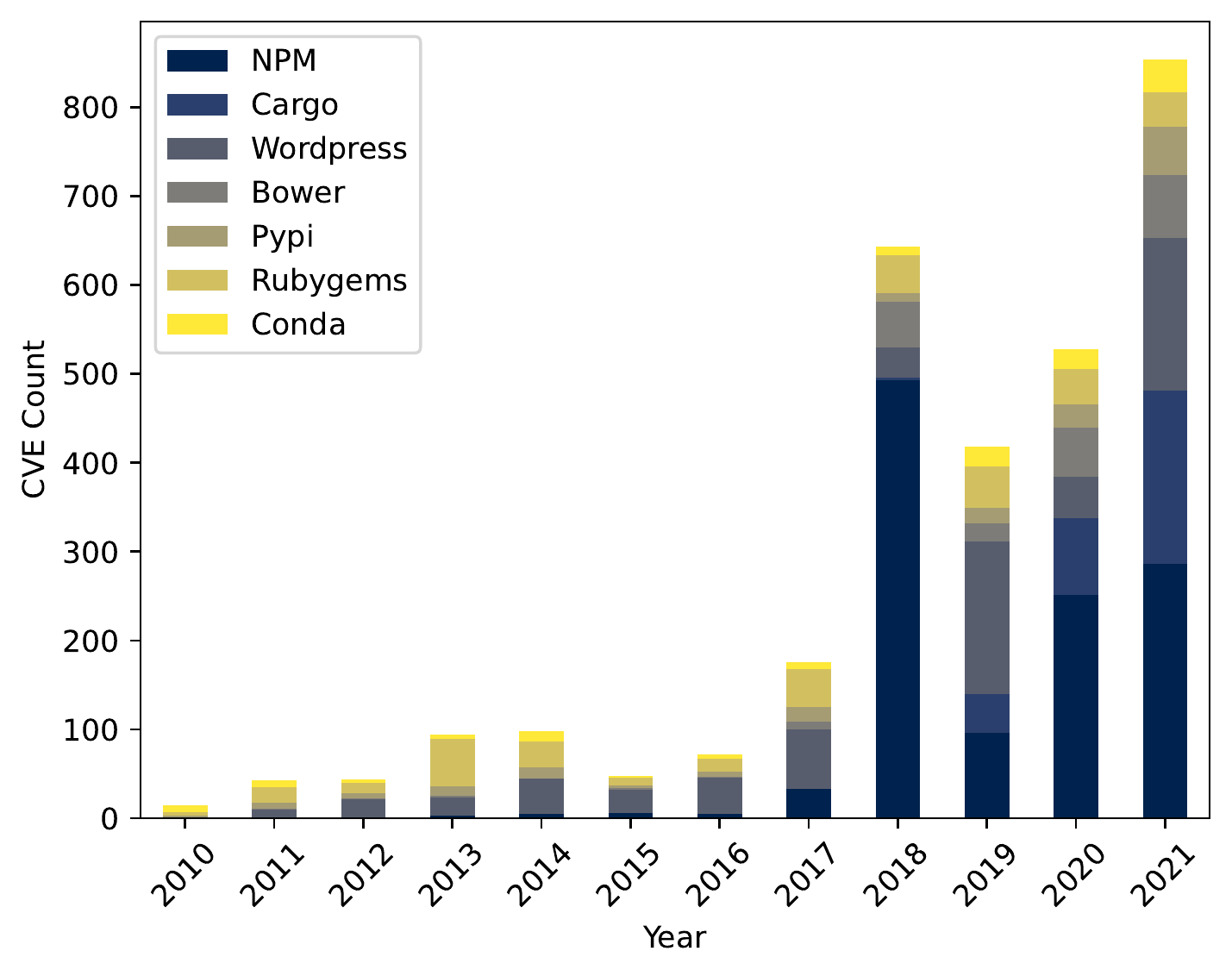}
    \caption{Comparison of CVE count per package manager per year including seven package managers containing most entries.\label{fig:cve_per_year}}
    \end{minipage}
\end{figure}


\paragraph{Discussion:} 
The results of this mapping, i.e. the vulnerable package counts, is compared with the other matching approaches in \Cref{fig:cve_per_pm}. The figure is limited to the top-10 package managers wrt. number of mapped packages.
The \textit{strict name mapping} approach yields an unexpected low number of mapped packages for certain package managers. We identified the following issues: (i) In particular, Go packages provide the repository identifier consisting of provider, owner and repository name in the Libraries.io dump, while the CVE entry contains only the repository name. (ii) The CPE strings of CVE entries contain potentially wrong attribute values or lack some information as for instance the \textit{target\_sw} value. (iii) Additionally, various entries do not specify the package manager, neither in the CPE string nor inside the \textsc{Summary}. 

Additionally to comparing the amount of vulnerable packages, we grouped the CVE entries per year and package manager if we were able to create a mapping with the \textit{strict name mapping} approach. This distribution as well as the trend of the top-7 package managers with the highest CVE count are depicted in \Cref{fig:cve_per_year}. Despite some fluctuating values, there is a noticeable upwards trend especially for the last few years. 


\subsubsection{Partial and Fuzzy Name Matching\label{ssec:partial}}
In our second approach we build a lookup table for each package manager: we define specific keywords (e.g., NPM, Maven, etc.) for the CVE \textsc{Summary} field and domain names or partial URLs (e.g., npmjs.com, pypi.org, etc.) for the \textsc{References}.
We iterate over each CVE entry, assign a package manager based on the lookup table, and check if the Libraries.io \textsc{Platform} value equals the assigned package manager and if a value of the CVE \textsc{Product} list occurs either in the \textsc{Name} or \textsc{Keywords} fields. 
In case multiple Libraries.io entries match the CVE entry, our program uses a fuzzy string matching method provided by the fuzzyset2 python package\footnote{\url{https://pypi.org/project/fuzzyset2}, last accessed 2023-02-09} with a \textit{cutoff} value of $0.3$ and chooses the entry with the highest similarity value. This low cutoff value was empirically chosen since we observed a number of projects with matching substrings but mismatching content length and surroundings.

\paragraph{Discussion:} 
The results of this approach are shown in \Cref{fig:cve_per_pm}. The vulnerable package counts of the \textit{partial and fuzzy name matching} are considerably lower than compared to the strict name mapping approach. Due to choosing the best result of the fuzzy string matching as well as the necessary \textit{cutoff} to avoid high false-positive rates, the approach is prone (i) to miss sub-packages with the same package name
and (ii) to miss matches because of the \textit{cutoff} value.


\subsubsection{Repository Matching}
\label{sub:repository matching}
In order to increase the accuracy of the matching of software packages to CVE entries, we focus in this approach on the references in the respective CVE entries.
The hypothesis is that the URLs in the references consist of the repository owner and repository name.\footnote{Consider for instance the GitHub URLs, which include the organisation/owner and repository.}


We use the value of the Libraries.io \textsc{Repository URL} field and extract the repository provider (i.e., github.com, bitbucket.org or gitlab.com) 
as well as the owner and the name of the repository. 
\textsc{repo\_link} contains the concatenated values of those fields. 
Additionally, we search for partial URLs of the format \textit{provider.tld/owner/repository} in the CVE \textsc{References} fields and store the results in a new array \textsc{links}.

We iterate over the CVE entries and check whether the values of the \textsc{links} array occur in the \textsc{repo\_link} field of any Libraries.io entry. 
First, we assigned the CVE entries to every package that matched the repository link resulting in very different vulnerable package counts compared to the other approaches. 
These results are shown in the ``All Links'' column of \Cref{tab:exp3_results}. To determine the amount of multiply assigned CVE entries, we reapplied the approach and only counted the first match of the repository link. Notably, in the ``First Link'' column of \Cref{tab:exp3_results}, the vulnerable package count is considerably lower. \Cref{fig:cve_per_pm} compares the ``First Link'' results to the other matching methods.

\begin{table}[!htb]
    \caption{The left table gives the top-5 vulnerable package count per package manager using the \textit{repository matching} approach. ``All Links'' counts all packages with a matching link, ``First Links'' counts the first match. The right table gives the 10 most occurring repositories in the Libraries.io dataset.\label{tab:vuln_results}}
    \begin{minipage}[t]{.4\linewidth}
      \centering
    \label{tab:exp3_results}
    \begin{tabular}{lrr}
    \toprule
    P.M. &  All Links &  First Link \\
    \midrule
    Go              &      13201 &         199 \\
    NPM             &       5359 &         510 \\
    Maven           &       3396 &         171 \\
    Pypi            &        458 &         224 \\
    Ruby            &        432 &         152 \\
    \bottomrule
    \end{tabular}
    \end{minipage}    
    \begin{minipage}[t]{.59\linewidth}
        \centering
    \label{tab:exp3_top_links}
    \begin{tabular}{lr}
    \toprule
    Repository &  Count \\
    \midrule
    github.com/openshift/origin          &   1523 \\
    github.com/kubernetes/kubernetes     &   1181 \\
    github.com/liferay/liferay-portal    &    508 \\
    github.com/facebook/create-react-app &    508 \\
    github.com/lodash/lodash             &    499 \\
    \bottomrule
    \end{tabular}
    \end{minipage}   
\end{table}

\paragraph{Discussion:}
We found, that various entries in the Libraries.io dataset contain the same \textsc{Repository URL} or at least the same combination of repository provider, owner and repository name. A lot of them seem to be sub-packages that are contained inside one repository or they erroneously contain the URL of the main project. \Cref{tab:exp3_top_links} shows the ten most common \textsc{repo\_link} values and the respective package count.


\subsection{Identified Challenges}
To summarize, we identified the following challenges when mapping CVE to open source packages:

\begin{enumerate}
    \item \textit{No clear repository identifiers:} The CVE entries do not contain clear identifiers of the code base in the package manager and/or software repository.
    \item \textit{Incomplete or wrong CVE entries:} The CVE list potentially contains wrong attribute values or lacks some critical information to map to code bases.
    \item \textit{Multiple relevant software packages:} In some cases, the CVE entry relates to multiple software packages and potentially multiple code repositories. 
    \item \textit{Reliable datasets:} There is no complete and reliable index of open source projects; the Libraries.io data is potentially erroneous and incomplete.
\end{enumerate}

The mapping approaches allowed us to derive some insights regarding the number of vulnerabilities for certain package managers (cf. \Cref{fig:cve_per_pm}), in particular, the strict name mapping approach displayed a noticeable upwards trend in reported vulnerabilities in recent years (\Cref{fig:cve_per_year}), however, an improved approach and a thorough evaluation is necessary to reach robust conclusions.

\section{Conclusion}
\label{sec:conclusion}


In this work, we have provided insights in the landscape of open-source projects in popular package managers, and have discussed three approaches to map an up-to-date list of CVE entries to their respective software package. 
In our analyses we have discussed both, shortcomings and quality issues of the available data, and shortcomings of the mapping wrt. accuracy and false-positive rate.



In future work, we plan the following research directions: Firstly, a thorough evaluation of the mapping approaches is required to provide more accurate results and to provide a large-scale mapping of CVE entries.
Second, performing data cleansing on the Libraries.io and the CVE dataset 
would highly increase accuracy and would enable further analyses about software packages and their vulnerabilities.
Third, research in improving the CVE standard; adding for instance additional meta-information such as a project identifier or project URL. 

Another possibility to improve our work is to consider alternative data sources, such as the GHTorrent\footnote{\url{gousios2014lean}} database -- a large and rich (meta)database of GitHub projects -- as an additional DB to map with.

\subsubsection{Acknowledgements}
This research was funded by the Austrian Research Promotion Agency (FFG) Bridge project 880592 ``SecDM -- Sichere und vertrauens-würdige on-premise data markets''. The financial support by the Austrian Research Promotion Agency is gratefully acknowledged.
We are grateful to the anonymous referees for suggesting numerous improvements.

\bibliographystyle{splncs04}
\bibliography{Bibliography}

\end{document}